\documentclass[twocolumn,aps,prl,floatfix,superscriptaddress,linenumbers]{revtex4}
\usepackage{tikz}
\usepackage{amsmath}
\usepackage[colorinlistoftodos, color=green!40, prependcaption]{todonotes}
\usepackage{graphicx,float}
\usepackage{times,txfonts}
\usepackage{textcomp}
\usepackage{multirow}
\usepackage{color}
\usepackage{soul}
\usepackage{hyperref}
\usepackage{nicefrac}
\usepackage{blindtext}
\usepackage{soul}
\usepackage{changes}

\usepackage[export]{adjustbox}
\usepackage{siunitx}
\usepackage[abbreviations]{glossaries-extra}
\usepackage{etoolbox}
\preto\section{\glsresetall}

\newcommand{\fref}[1]{Fig.~\ref{#1}}

\glssetcategoryattribute{abbreviation}{nohyper}{true}
\makeglossaries
\newabbreviation{eo}{EO}{electro-optic}
\newabbreviation{mzm}{MZM}{Mach-Zehnder modulator}
\newabbreviation{cpw}{CPW}{coplanar waveguide}
\newabbreviation{gsg}{GSG}{ground-signal-ground}
\newabbreviation{solt}{SOLT}{short-open-load-through}
\newabbreviation{imdd}{IMDD}{intensity-modulation and direct-detection}
\newabbreviation{dsp}{DSP}{digital signal processing}
\newabbreviation{snr}{SNR}{signal-to-noise ratio}
\newabbreviation{ltsi}{LT-on-Si}{lithium-tantalate-on-silicon}
\newabbreviation{ltoi}{LTOI}{lithium-tantalate-on-insulator}

\begin{document}
	
	\title{Broadband suspended lithium tantalate Mach–Zehnder modulator achieving a 460~Gbit/s net data rate}

	\author{Zihan Li}
	\thanks{Equal contribution for this work}
	\affiliation{Institute of Physics, Swiss Federal Institute of Technology, Lausanne (EPFL), CH-1015 Lausanne, Switzerland}
	\affiliation{Institute of Electrical and Micro Engineering (IEM), EPFL, CH-1015 Lausanne, Switzerland}
	
	\author{Alexander Kotz}
	\thanks{Equal contribution for this work}
	\affiliation{Institute of Photonics and Quantum Electronics (IPQ), Karlsruhe Institute of Technology (KIT), 76131 Karlsruhe, Germany}
	
	\author{Adrian Schwarzenberger}
	\affiliation{Institute of Photonics and Quantum Electronics (IPQ), Karlsruhe Institute of Technology (KIT), 76131 Karlsruhe, Germany}
	
	\author{Christian Koos}
	\affiliation{Institute of Photonics and Quantum Electronics (IPQ), Karlsruhe Institute of Technology (KIT), 76131 Karlsruhe, Germany}
	
	\author{Tobias J. Kippenberg}
	\email{tobias.kippenberg@epfl.ch}
	\affiliation{Institute of Physics, Swiss Federal Institute of Technology, Lausanne (EPFL), CH-1015 Lausanne, Switzerland}
	\affiliation{Institute of Electrical and Micro Engineering (IEM), EPFL, CH-1015 Lausanne, Switzerland}

	\begin{abstract}
		Thin-film lithium tantalate \textrm{LiTaO$_3$} photonic integrated circuits have recently been demonstrated as a promising next-generation electro-optic platform, offering favorable properties including reduced DC drift, higher optical power handling, and lower birefringence compared to lithium niobate.
		However, high-speed \textrm{LiTaO$_3$} modulators reported to date have predominantly relied on silicon substrates, whose large dielectric constant ($\varepsilon_r \approx 11.7$) compromises microwave velocity matching and imposes RF conductor losses that limit the achievable electro-optic bandwidth. 
		Here, we implement a silicon substrate undercut technique to suspend the electrode region of lithium-tantalate-on-insulator (LTOI) Mach-Zehnder modulators (MZMs), effectively decoupling the traveling-wave electrodes from the high-permittivity silicon handle wafer, thereby reducing microwave losses. In addition, the undercut removes any susceptibility to parasitic surface conductance (PSC) induced losses of the oxide-silicon interface.
		The fabricated MZM achieves a \SI{3}{dB} electro-optic bandwidth of \SI{110}{\giga\hertz}, with a half-wave voltage of \SI{5.1}{\volt} ($V_\pi L = \SI{4}{\volt\centi\meter}$) for an \SI{8}{\milli\meter}-long device.
		Exploiting the extended bandwidth, we demonstrate a high single-lane intensity-modulation and direct-detection (IMDD) net data rate of \SI{460}{\giga\bit\per\second}~using PAM8 signaling.
		These results establish silicon substrate undercut as an effective and process-compatible pathway to unlock the full electro-optic potential of lithium tantalate on its native silicon-based wafer platform.
	\end{abstract}
	
	\maketitle
	
	\vspace{5mm}

	\section{Introduction}
	
	The explosive growth of artificial intelligence workloads and hyper-scale data-center deployments is driving urgent demand for optical interconnect at all scales, from short range (within data centers), to long range (between the data centers), with high data rates and lower energy consumption~\cite{LightmatterOFC2025}. Photonic integrated circuit-based electro-optic modulators based on thin-film ferroelectric materials, such as lithium niobate ($\mathrm{LiNbO_3}$) and lithium tantalate ($\mathrm{LiTaO_3}$), have attracted considerable attention as they simultaneously offer large electro-optic coefficients, low drive voltages, and broad modulation bandwidths well suited for single-lane rates beyond \SI{400}{\giga\bit\per\second}~\cite{Wang:18,Berikaa_TFLN}. This makes this platform in particular promising for coherent communications, that utilize complex modulation formats, and is used in long-haul optical communications. Similarly, it has been considered recently for linear drive pluggable optics (LPO)~\cite{st2026driver,st2024practical}.
	
	Recently, low loss lithium tantalate photonic integrated circuits have been demonstrated as an alternative ferroelectric material platform~\cite{WangNature:24,WangOptica:24}.	Lithium tantalate on insulator (LTOI) is already today commercially deployed in RF surface acoustic wave (SAW) filters for RF front ends and therefore benefits from economies of scales i.e. low costs substrates driven by volume applications, which is not the case for lithium niobate on insulator (LNOI).
	From a materials perspective compared with \textrm{LiNbO$_3$}, \textrm{LiTaO$_3$} modulators exhibit improved DC bias stability~\cite{lin2026copper,powell2024dc,sayem2026high}, greater tolerance to high optical power with modulators up to \SI{1.17}{\watt} demonstrated~\cite{lin2026copper}, and markedly lower birefringence ($\delta n = 0.004$ vs.\ $\delta n = 0.074$ for \textrm{LiNbO$_3$} at \SI{1550}{\nano\meter}), simplifying the design of complex photonic circuits such as arrayed waveguide gratings (AWG)~\cite{Hulyal:25}, which are key building blocks for wavelength division multiplexing components - as well as enabling broadband electro-optic combs~\cite{zhang2025ultrabroadband}. Moreover, the platform has been combined with the copper Damascene process, enabling direct copper-to-copper bonding with electronic driver integrated circuits (IC)~\cite{lin2026copper}.
	Electro-optic bandwidths exceeding \SI{110}{\giga\hertz} have been demonstrated for $\mathrm{LiNbO_3}$~modulators on silicon substrates~\cite{WangOptica:24}; however, the high relative permittivity of silicon ($\varepsilon_r \approx 11.7$) introduces an inherent mismatch between the microwave effective index and the optical group index. 
	Achieving simultaneous impedance and velocity matching under these conditions forces a narrow signal conductor geometry, which in turn increases RF conductor losses~\cite{li2026low}~and limits the practical electrode length to around \SI{6}{\milli\meter}, yielding relatively large half-wave voltages of approximately \SI{4.8}{\volt}~\cite{WangOptica:24}.
	
	Several strategies have been explored to overcome the dielectric substrate bottleneck.
	Replacing the silicon carrier with a low-permittivity material such as fused silica~\cite{Kharel:21} removes the velocity-matching constraint and allows a wider electrode geometry, substantially reducing microwave attenuation and extending the interaction length.
	
	An alternative route that avoids the complexity of full wafer bonding is to selectively remove the silicon beneath the electrode region through a substrate undercut process~\cite{chen2022high}.
	This approach retains the standard LTOI starting wafer and wafer-scale-compatible processing, while effectively suspending the traveling-wave electrodes in air, which has the lowest-permittivity and lowest dielectric loss, between the probe contact pads. The suspended membrane decouples the RF mode from the silicon handle wafer, enabling velocity matching with a wider electrode geometry and drastically reduced microwave loss.
		Here, we implement a silicon undercut process for LTOI~Mach-Zehnder modulators (MZMs) and demonstrate that the suspended electrode architecture provides a significant improvement in electro-optic bandwidth for a device with an \SI{8}{\milli\meter}-long modulation section.
	We further validate the high-speed performance through \gls{imdd} signaling experiments, achieving a net data rate of \SI{460}{\giga\bit\per\second} for the $\mathrm{LiNbO_3}$~platform.
	Our results demonstrate that the silicon undercut offers an accessible, wafer-compatible route to high-performance LTOI modulators without the need for a dedicated low-permittivity substrate.

	\begin{figure*}[t]
		\centering
		\includegraphics[width=\linewidth]{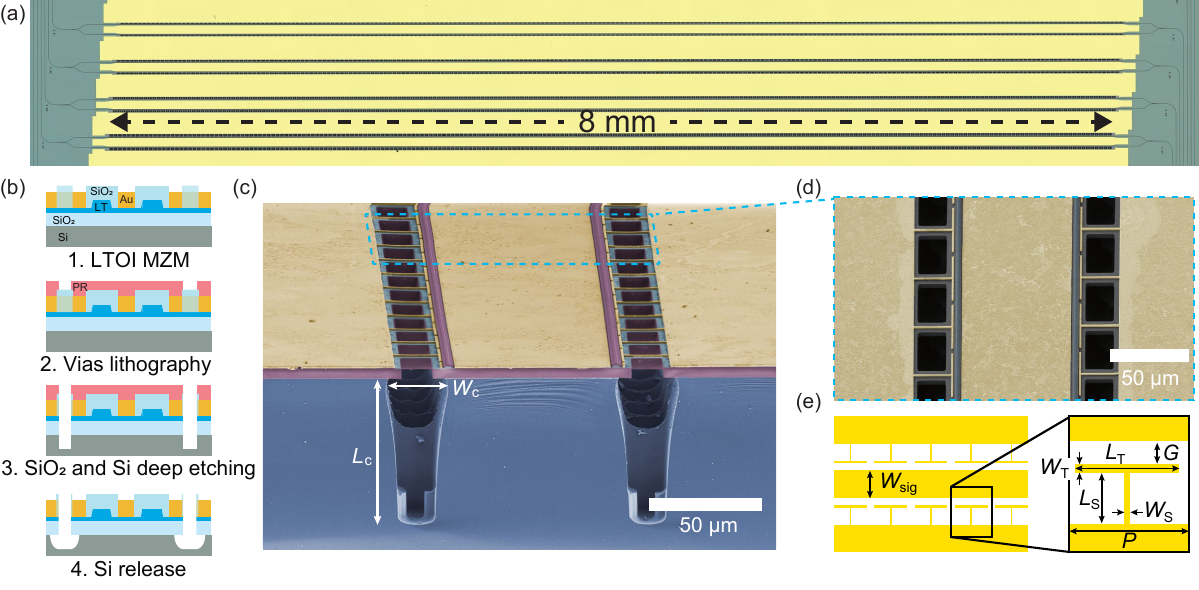}
		
		\caption{\textbf{Suspended thin-film lithium tantalate Mach-Zehnder modulator fabricated via silicon substrate undercut.}
			\textbf{(a)} Optical microscope image of a fabricated chip showing the \SI{8}{\milli\meter}-long modulator electrode region.
			\textbf{(b)} Schematic cross-sectional process flow illustrating the four key fabrication stages:
			1.~Starting LTOI \gls{mzm} on Si with Au electrodes and SiO$_2$ cladding;
			2.~Via lithography to define the release holes;
			3.~Deep SiO$_2$ and Si dry etching through the vias;
			4.~Isotropic Si release to suspend the modulation region, leaving a free-standing membrane comprising the \textrm{LiTaO$_3$} waveguide, SiO$_2$ cladding, and Au electrodes. 
			Cross-section \textbf{(c)}~and top view \textbf{(d)}~scanning electron microscope (SEM) pictures of the modulation region after silicon release confirms successful suspension of the electrode stack. The undercut cavities beneath the ground electrodes are clearly visible.
			\textbf{(e)} Modelled suspended CPW geometry, showing the T-shaped ground-electrode segments and the key design parameters ($W_\mathrm{sig}$, $P$, $G$, $L_S$, $W_S$, $L_T$, $W_T$). 
		}
		\label{fig:fig1}
	\end{figure*}

	\section{Results}
	
	\subsection{Device Design and Fabrication}
	
	The fabricated device, shown in \fref{fig:fig1}~(a), consists of two \SI{8}{\milli\meter}-long suspended electro-optic phase shifters combined with 1~$\times$~2 multimode-interference (MMI) couplers to form a push-pull Mach-Zehnder modulator.

	The suspended LTOI MZMs are built on a commercial 4-inch X-cut LTOI wafer (NANOLN), comprising a \SI{600}{\nano\meter}-thick $\mathrm{LiTaO_3}$ single-crystal thin film, a \SI{4.7}{\micro\meter}-thick SiO$_2$ buried oxide layer, and a \SI{525}{\micro\meter}-thick high-resistivity silicon substrate.
	In contrast to approaches requiring a thick silicon dioxide insulating layer~\cite{xue2026high} or a full silicon dioxide substrate~\cite{Kharel:21}, the silicon undercut strategy adopted here selectively removes the silicon only beneath the electrode region, preserving the mechanical integrity of the chip while eliminating the dielectric loading of the traveling-wave \gls{cpw}.
	
	The optical waveguides are patterned using deep-ultraviolet (DUV) stepper lithography followed by argon ion beam etching with a diamond-like-carbon hard mask, yielding an etch depth of \SI{400}{\nano\meter} with a \SI{200}{\nano\meter}-thick residual slab, consistent with established LTOI fabrication processes~\cite{Li:23,WangNature:24}.
	The modulator arms are clad with a \SI{1.5}{\micro\meter}-thick silicon dioxide layer deposited by inductively coupled plasma chemical vapor deposition (ICP CVD)~\cite{Qiu:2024}. 
	Traveling-wave gold electrodes (\SI{15}{\nano\meter} Ti / \SI{800}{\nano\meter} Au) are then defined by DUV lithography, dry etching, metal evaporation, and lift-off, following the same process flow as for our previous LTOI devices~\cite{WangNature:24}.
	
	The silicon undercut is carried out as a post-electrode processing step, as illustrated in \fref{fig:fig1}~(b).
	First, via patterns are defined by ultraviolet (UV) lithography in-between the capacitively loaded CPW (\fref{fig:fig1}~(d)). Then trenches with a depth ($L_c$) of \SI{80}{\micro\meter} are anisotropically etched through the silicon dioxide and into the silicon substrate using a Bosch deep reactive-ion etching process. To optimize the silicon undercut width for proper velocity matching, the chips are separated from the wafer by silicon deep etching and backside grinding for individual process\cite{liu2021high}. Owing to the chip singulation process and the inverse taper edge coupler, a typical fiber-to-fiber coupling loss of \SI{-5.2}{\decibel} is achieved.
	Subsequently, an isotropic silicon etch selectively removes the silicon beneath the electrode region through the vias, leaving the $\mathrm{LiTaO_3}$~photonic layer surrounded by silicon dioxide and the Au electrodes suspended above an air-filled cavity (\fref{fig:fig1}~(c)).
	The undercut width $W_c$ is experimentally optimized on each chip to achieve phase matching between the microwave and optical signals.
	The suspended membrane extends along the full \SI{8}{\milli\meter} modulation section of the modulator.
	
	To preserve the mechanical robustness of the fragile suspended modulation arms, the T-shaped undercut segments are applied only to the ground electrodes, while the silicon dioxide layer beneath the signal electrode remains continuous after the silicon undercut.
	Furthermore, combining anisotropic and isotropic silicon etching reduces the proportion of the microwave field in the residual silicon, enabling phase matching with a smaller lateral undercut extent and thereby improving mechanical stability.
	Accounting for fabrication tolerances, impedance matching, and phase matching, the CPW design parameters are chosen as ($W_\mathrm{sig}$, $P$, $G$, $L_S$, $W_S$, $L_T$, $W_T$) = (70, 35, 6, 20, 1.5, 33, 1.5)~$\mu$m (\fref{fig:fig1}~(e)).
	The electrode cross-section above the air gap closely resembles that of a modulator on a low-permittivity substrate, substantially relaxing the trade-off between impedance matching and conductor width that restricts designs on unsuspended silicon.

	\begin{figure*}[t]
		\centering
		\includegraphics[width=0.9\linewidth]{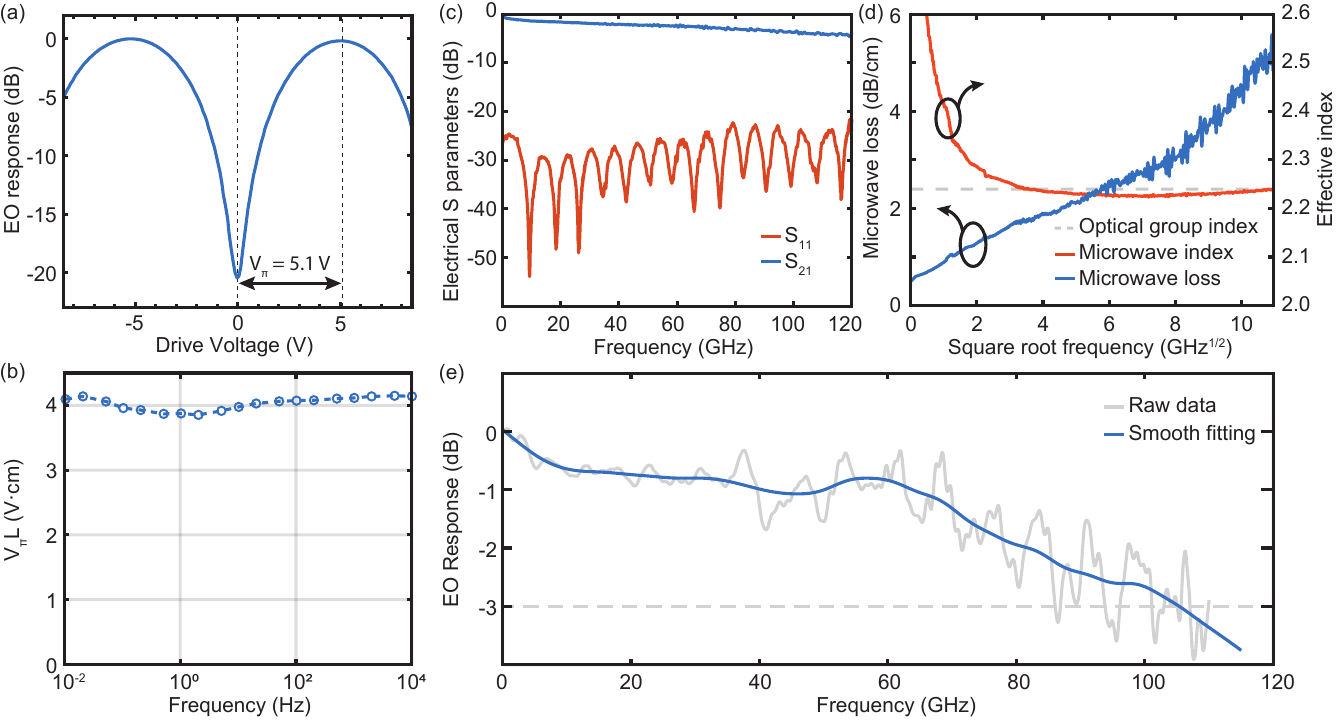}
		\caption{\textbf{Modulation performance characterization of the suspended $\mathrm{LiTaO_3}$ Mach-Zehnder modulator.}
			\textbf{(a)} Normalized optical transmission as a function of applied voltage at a carrier wavelength of \SI{1550}{\nano\meter}, measured at \SI{100}{\hertz} with a sawtooth waveform. A half-wave voltage of $V_\pi = \SI{5.1}{\volt}$ is extracted from the fitted curve.
			Measured \textbf{(b)} $V_\pi L$ and \textbf{(c)} normalized EO response as a function of modulation frequency from \SI{10}{\milli\hertz} to \SI{10}{\kilo\hertz}, confirming DC bias stability with negligible drift across four decades of frequency.
			\textbf{(d)} Electrical S-parameters ($S_{11}$, orange; $S_{21}$, blue) of the suspended traveling-wave \gls{cpw} up to \SI{120}{\giga\hertz}.
			\textbf{(e)} Calculated microwave attenuation (blue) and effective microwave phase index (orange) of the \gls{cpw} as a function of the square root of frequency. The microwave effective index closely matches the simulated optical group index (dashed), confirming velocity matching in the high-frequency region.
			\textbf{(f)} Measured EO response of the suspended \gls{mzm}, normalized to the value at \SI{1}{\giga\hertz}. Raw data (light gray) and a smoothed fit (blue) are shown. A \SI{3}{dB} bandwidth of \SI{110}{\giga\hertz} is achieved.}
		\label{fig:fig2}
	\end{figure*}

	\subsection{Electro-optic Modulation Efficiency}
	
	The modulation efficiency of the fabricated suspended \gls{mzm} is quantified through the half-wave voltage--length product ($V_\pi L$).
	A low-frequency sawtooth voltage waveform with a peak-to-peak amplitude of \SI{20}{\volt} is applied to the modulator electrodes, and the normalized optical transmission is recorded simultaneously using a photodetector (DET08CFC) and an oscilloscope (Rohde\,\&\,Schwarz RTA4004).
	The half-wave voltage is extracted by fitting the measured electro-optic response curve.
	
	As summarized in \fref{fig:fig2}~(a), the \SI{8}{\milli\meter}-long suspended \gls{mzm} exhibits a half-wave voltage of \SI{5.1}{\volt} at a carrier wavelength of \SI{1550}{\nano\meter}, corresponding to a $V_\pi L$ product of \SI{4}{\volt\centi\meter}.
	The DC bias stability of the device is verified by measuring $V_\pi L$ as a function of modulation frequency from \SI{10}{\milli\hertz} to \SI{10}{\kilo\hertz}, with no appreciable drift observed across this four-decade range, consistent with the improved DC stability characteristic of the \textrm{LiTaO$_3$} platform compared to the \textrm{LiNbO$_3$} platform~\cite{Holzgrafe:24}.

	\subsection{High-Frequency Characterization}
	
	The high-frequency electrical and electro-optic behaviour of the suspended \gls{mzm} is characterized using a vector network analyzer (VNA; ME7838AX, Anritsu Corporation) operating from \SI{70}{\kilo\hertz} to \SI{125}{\giga\hertz}, with \SI{110}{\giga\hertz} \gls{gsg} RF probes (T110A-GSG0100, MPI Corporation).
	A two-port \gls{solt} calibration on a commercial substrate (AC2-2, MPI Corporation) is used to set the measurement reference planes at the probe tips.
	
	Figure~\ref{fig:fig2}~(d) shows the electrical S-parameters of the suspended \gls{cpw}.
	The reflection coefficient $S_{11}$ remains below \SI{-20}{\decibel} across the entire measured frequency range, indicating good impedance matching to the \SI{50}{\ohm} source.
	The transmission coefficient $S_{21}$ exhibits a gradual roll-off, reaching \SI{-4.3}{\decibel} at \SI{120}{\giga\hertz}.
	The microwave attenuation, plotted against the square root of frequency in \fref{fig:fig2}~(e), follows a near-linear trend, confirming that conductor (skin-depth) loss is the dominant dissipation mechanism~\cite{simons2004coplanar}.
	The extracted effective microwave phase index is in close agreement with the simulated optical group index of the \textrm{LiTaO$_3$} waveguide at \SI{1550}{\nano\meter}, confirming velocity matching in the high-frequency region (\fref{fig:fig2}~(d)).
	
	To measure the electro-optic frequency response, an optical carrier at \SI{1550}{\nano\meter} is coupled into the modulator.
	The device is driven from the first VNA port, while the second port is connected to a high-speed photodiode that detects the modulated optical signal.
	The output port of the \gls{cpw} is terminated with a \SI{50}{\ohm} resistor via a second RF probe to suppress back-reflections.
	De-embedding is applied to shift the measurement reference planes to the input probe tip and the optical output of the MZM, using the frequency responses of the probe and photodiode supplied by their respective manufacturers.
	
	The resulting electro-optic response, shown in \fref{fig:fig2}~(f), is normalized to the value at \SI{1}{\giga\hertz}.
	The smoothed curve yields a \SI{3}{dB} bandwidth of \SI{110}{\giga\hertz}.
	Compared to previously reported LTOI modulators on unsuspended silicon substrates with comparable active lengths~\cite{WangOptica:24}, the undercut architecture provides a substantial bandwidth improvement, demonstrating that selective substrate removal is an effective strategy to mitigate the velocity-mismatch and conductor-loss penalties imposed by the high-permittivity silicon carrier, as well as the potential parasitic surface-conductance loss in the silicon~\cite{lederer2003substrate}.

	\begin{figure*}[t]
		\centering
		\includegraphics[width=0.9\linewidth]{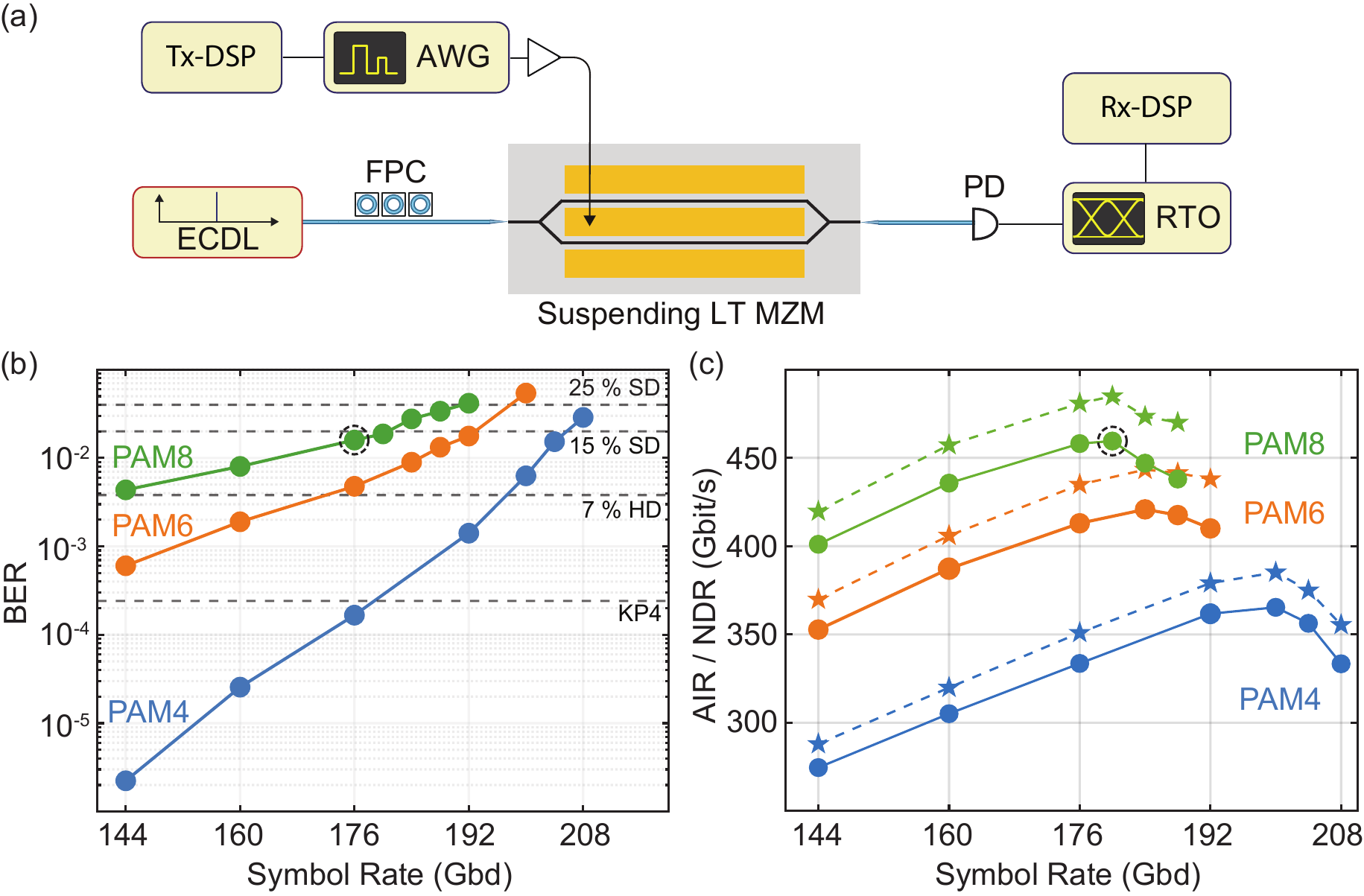}
		\caption{\textbf{Intensity-modulation and direct-detection (IMDD) signaling experiment using the suspended lithium-tantalate-on-insulator Mach-Zehnder modulator.}
			\textbf{(a)} Experimental setup. An external-cavity diode laser (ECDL) provides the optical carrier, and a fiber polarization controller (FPC) adjusts the input polarization. The electrical drive signal is synthesized by transmitter digital signal processing (Tx-DSP) and generated by an arbitrary waveform generator (AWG). The modulated signal is detected by a high-speed photodiode (PD) connected to a real-time oscilloscope (RTO), followed by offline receiver DSP (Rx-DSP). The chip under test is the suspended $\mathrm{LiTaO_3}$~\gls{mzm}.
			\textbf{(b)} Measured bit error ratios (BER) as a function of symbol rate for PAM4 (blue), PAM6 (orange), and PAM8 (green) signals. Horizontal dashed lines indicate the \SI{25}{\percent} and \SI{15}{\percent} soft-decision forward error correction (SD-FEC) thresholds and the \SI{7}{\percent} hard-decision FEC (HD-FEC) threshold as well as the threshold for KP4.. The circled marker indicates the operating point achieving the highest net data rate of \SI{460}{\giga\bit\per\second}.
			\textbf{(c)} Achievable information rates (AIR, dashed lines) and corresponding net data rates (NDR, solid lines) as a function of symbol rate for PAM4 (blue), PAM6 (orange), and PAM8 (green).}
		\label{fig:fig4}
	\end{figure*}

	\subsection{High-Speed IMDD Signaling with Suspended Lithium Tantalate MZM}
	To evaluate the system-level performance of the suspended lithium-tantalate-on-insulator MZM, we conduct high-speed \gls{imdd} signaling experiments with the configuration depicted in \fref{fig:fig4}~(a).
	An external-cavity diode laser (ECDL) operating at \SI{1550}{\nano\meter} provides the optical carrier with an output power of \SI{17.8}{dBm}.
	The polarization is aligned to the quasi-transverse-electric mode of the $\mathrm{LiTaO_3}$~waveguide using a fiber polarization controller (FPC), and lensed fibers are employed for chip-to-fiber coupling.
	
	Electrical drive signals are generated by a high-speed arbitrary waveform generator (AWG; M8199B, Keysight Technologies Inc.) operating at a sampling rate of \SI{256}{\giga Sa\per\second}, routed through an RF cable and a broadband amplifier (Amp.; AH15199B, Anritsu Corporation), and applied to the \gls{cpw} via a \SI{110}{\giga\hertz} RF probe.
	Pulse-amplitude modulated (PAM) waveforms with four, six, and eight levels are generated from pseudo-random bit sequences with pattern lengths exceeding $2^{17}$ bits using offline Tx-DSP, which includes root-raised cosine pulse shaping with a roll-off factor of $\beta = 0.05$, predistortion based on linear minimum-mean-square-error equalization to compensate for the transmitter frequency response excluding the LTOI MZM, and signal clipping to a peak-to-average power ratio of \SI{9}{\decibel}.
	The output \gls{cpw} is terminated with a \SI{50}{\ohm} resistor through a second \SI{110}{\giga\hertz} RF probe and a broadband bias-tee, which simultaneously sets the quadrature operating point of the modulator.
	The received optical signal is digitized by a real-time oscilloscope (RTO; UXR~1004A, Keysight Technologies Inc.) with a \SI{256}{\giga Sa\per\second} sampling rate and \SI{104}{\giga\hertz} analog bandwidth, after direct detection by a high-speed photodiode (PD; Fraunhofer HHI).
	Receiver-side offline DSP includes resampling to two samples per symbol, timing recovery, Sato equalization, and additional least mean square-based adaptive equalization.
	
	Signaling experiments are performed at symbol rates ranging from \SI{144}{GBd} to \SI{208}{GBd} for PAM4, PAM6, and PAM8.
	The measured BER curves as a function of symbol rate are presented in \fref{fig:fig4}~(b), together with reference threshold lines for common forward error correction (FEC), i.e. KP4, hard-decision FEC with \SI{7}{\percent} overhead (\SI{7}{\percent} HD)~\cite{itu_g.975.1}, as well as soft-decision FEC with \SI{15}{\percent} (\SI{15}{\percent} SD) and \SI{25}{\percent} (\SI{25}{\percent} SD) overhead~\cite{GraelliAmat2020}.
	The BER remains below the \SI{25}{\percent} SD-FEC limit up to \SI{188}{GBd} for PAM8, up to \SI{192}{GBd} for PAM6 and up to \SI{208}{GBd} for PAM4 signaling. For the latter, the BER remain below the KP4 limit up to \SI{176}{GBd}.
	The achievable information rate (AIR) is computed via the generalized mutual information (GMI) assuming an additive white Gaussian noise channel model~\cite{gmi_ivanov}, and the net data rate (NDR) is derived by selecting appropriate FEC codes from the normalized GMI following~\cite[Table~2]{Hu2022fec}.
	The results are shown in \fref{fig:fig4}~(c).
	The highest AIR of \SI{485}{\giga\bit\per\second} is obtained for PAM8 at \SI{180}{GBd}.
	The maximum NDR of \SI{460}{\giga\bit\per\second} is achieved for PAM8 at a symbol rate of \SI{180}{GBd} using SD-FEC, establishing a high single-lane data rate for the lithium tantalate modulator platform and demonstrating performance on par with state-of-the-art thin-film lithium niobate modulators~\cite{Berikaa_TFLN}.

	\section{Conclusion}
	
	We have demonstrated a silicon substrate undercut approach to fabricate suspended thin-film lithium tantalate Mach-Zehnder modulators.
	By selectively removing the high-permittivity silicon beneath the traveling-wave electrodes through via-defined deep silicon etching followed by isotropic silicon release, the \gls{cpw} is effectively suspended in air, decoupling the microwave mode from the silicon handle wafer.
	This substantially reduces the microwave conductor losses with a significantly wider electrode geometry and relaxes the buried silicon dioxide thickness constraint for velocity-matching compared to conventional LTOI devices on unsuspended silicon.
	
	The fabricated \SI{8}{\milli\meter}-long suspended \gls{mzm} achieves a \SI{3}{dB} electro-optic bandwidth of \SI{110}{\giga\hertz} with a half-wave voltage of \SI{5.1}{\volt}, representing a marked improvement over non-suspended LTOI modulators of comparable length.
	The DC bias stability characteristic of the $\mathrm{LiTaO_3}$~material is fully preserved after substrate release, as confirmed by $V_\pi L$ measurements spanning more than four decades in modulation frequency with no appreciable drift.
	Leveraging the enhanced electro-optic bandwidth, we achieve a single-lane net data rate of \SI{460}{\giga\bit\per\second} for the $\mathrm{LiTaO_3}$~platform using PAM8 signaling in the C-band, on par with the best results reported for thin-film lithium niobate modulators.
	
	These results confirm that silicon substrate undercut is a viable and process-compatible route to high-performance $\mathrm{LiTaO_3}$~electro-optic modulators, without requiring wafer bonding to an alternative substrate.
	The approach is fully compatible with existing LTOI foundry workflows and can enable manufacturing of electro-optical modulators for next-generation coherent communications transceivers or linear drive pluggable optics within data-centers, as well as modulators for 5G/6G radio-over-fiber \cite{ren2019integrated}.

	\vspace{1em}
	
	\noindent{\textbf{Data and Code Availability.}}
	Data and code used to produce the figures within this paper will be available at \texttt{Zenodo} upon publication of the manuscript.
	
	\vspace{1em}
	
	\noindent\textbf{Funding.}

This work was supported by the Horizon Europe EIC Transition programme under grant agreement No. grant agreement 101131069 (ELLIPTIC) and under grant No. 101113260 (HDLN), as well as funding from the Swiss State Secretariat for Education, Research and Innovation (SERI). This work has received funding from the European Research Council (ERC) under the Horizon Europe research and innovation programme, grant agreement No. 101167540 (ATHENS), as well as from the German Research Foundation via the projects PACE (No. 403188360) and GOSPEL (No. 403187440).

	\vspace{1em}
	
	\noindent\textbf{Acknowledgements.}
	We acknowledge the EPFL Center of MicroNano Technology (CMi) and the Institute of Physics (IPHYS) cleanroom for supporting sample fabrication.
	
	\vspace{1em}
	
	\noindent\textbf{Competing Interests.}
	C.K.\ and T.J.K.\ are co-founders and shareholders of Luxtelligence SA, St.~Sulpice, Switzerland, a company engaged in electro-optic modulators based on ferroelectric materials, such as lithium niobate and lithium tantalate.
	The other authors declare no competing interests.
	
	\vspace{5mm}
	
	\bibliography{MZM.bib}
	
\end{document}